\documentclass[twocolumn,trackchanges]{aastex701}

\usepackage{csquotes}

\newcommand{\lmo}{LMC X-1}
\newcommand{\lmt}{LMC X-3}

\newcommand{\cyg}{Cyg X-1}

\newcommand{\gx}{GX 339-4}
\newcommand{\xto}{XTE J1859+226}
\newcommand{\xtt}{XTE J1650-500}
\newcommand{\bht}{4U 1630-47}
\newcommand{\bhi}{4U 1820-30}
\newcommand{\gxo}{GX 340+0}
\newcommand{\ax}{AX 1745.6-2901}
\newcommand{\xlcp}{{\it XL-Calibur}}
\newcommand{\nic}{{\it NICER}}
\newcommand{\nus}{{\it NuSTAR}}
\newcommand{\maxi}{{\it MAXI}}
\newcommand{\sw}{{\it Swift}}
\newcommand{\bat}{{\it BAT}}
\newcommand{\ixpe}{{\it IXPE}}
\newcommand{\intg}{{\it INTEGRAL}}

\newcommand{\rxt}{{\it RXTE}}
\newcommand{\pgp}{{\it PoGO+}}
\newcommand{\suz}{{\it Suzaku}}

\begin{document}

\title[\cyg\ variability]{Is the hard X-ray polarization of \cyg{} stable?}

\correspondingauthor{*Varun (vvarun@kth.se)}

\author[0000-0003-4433-1365]{Matthew G. Baring} \affiliation{Department of Physics and Astronomy -- MS 108, Rice University, 6100 Main Street, Houston, Texas 77251-1892, USA} \email{baring@rice.edu}

\author[0009-0009-3051-6570]{Jacob Casey} \affiliation{Department of Physics and Astronomy and Space Science Center, University of New Hampshire, 8 College Road, Durham, NH 03824, USA} \email{jacob.j.casey@unh.edu}

\author[0000-0002-5250-2710]{Ephraim Gau} \affiliation{Department of Physics, Washington University, 1 Brookings Drive, Saint Louis, MO 63130, USA} \email{ephraimgau@wustl.edu}

\author[0000-0002-0987-0278]{Tomohiro Hakamata} \affiliation{Department of Earth and Space Science, The University of Osaka, 1-1 Machikaneyama-cho, Toyonaka, Osaka 560-0043, Japan} \email{hakamata@ess.sci.osaka-u.ac.jp}

\author[0000-0002-9705-7948]{Kun Hu} \affiliation{Department of Physics, Washington University, 1 Brookings Drive, Saint Louis, MO 63130, USA} \email{hkun@wustl.edu}

\author[]{Daiki Ishi} \affiliation{Faculty of Fundamental Science and Engineering, Chuo University, 1-13-27 Kasuga, Bunkyo-ku, Tokyo 112-8551, Japan} \email{dishi494@g.chuo-u.ac.jp}

\author[0000-0001-7477-0380]{Fabian Kislat} \affiliation{Department of Physics and Astronomy and Space Science Center, University of New Hampshire, 8 College Road, Durham, NH 03824, USA} \email{fabian.kislat@unh.edu}

\author[0000-0001-5191-9306]{Mózsi Kiss} \affiliation{KTH Royal Institute of Technology, Department of Physics, 106 91 Stockholm, Sweden} \affiliation{The Oskar Klein Centre for Cosmoparticle Physics, AlbaNova University Center, 106 91 Stockholm, Sweden}\email{mozsi@kth.se}

\author[0000-0002-1084-6507]{Henric Krawczynski} \affiliation{Physics Department and the McDonnell Center for the Space Sciences, Washington University in St. Louis, 1 Brookings Drive, Saint Louis, MO 63130, USA} \email{krawcz@wustl.edu}

\author[0009-0007-7463-5147]{Maitreya Kundu} \affiliation{Physics Department and the McDonnell Center for the Space Sciences, Washington University in St. Louis, 1 Brookings Drive, Saint Louis, MO 63130, USA} \email{k.maitreya@wustl.edu}

\author[0009-0006-9014-2716]{Haruki Kuramoto} \affiliation{Department of Earth and Space Science, The University of Osaka, 1-1 Machikaneyama-cho, Toyonaka, Osaka 560-0043, Japan} \email{kuramoto@ess.sci.osaka-u.ac.jp}

\author[0000-0002-5202-1642]{Lindsey Lisalda} \affiliation{Washington University in St. Louis} \email{Lindsey.Lisalda@wustl.edu}

\author[]{Masaki Maekawa}  \affiliation{Graduate School of Advanced Science and Engineering, Hiroshima University,  1-3-1 Kagamiyama, Higashi-Hiroshima, Hiroshima 739-8526, Japan} \email{maekawa@astro.hiroshima-u.ac.jp}

\author[]{Hironori Matsumoto} \affiliation{Department of Earth and Space Science, The University of Osaka, 1-1 Machikaneyama-cho, Toyonaka, Osaka 560-0043, Japan} \email{matumoto@ess.sci.osaka-u.ac.jp}

\author[0009-0005-0818-7484]{Shravan Vengalil Menon} \affiliation{Washington University in St. Louis} \email{s.vengalilmenon@wustl.edu}

\author[0009-0008-4216-064X]{Asca Miyamoto} \affiliation{Graduate School of Advanced Science and Engineering, Hiroshima University, 1-3-1 Kagamiyama, Higashi-Hiroshima, Hiroshima 739-8526, Japan} \email{miyamoto.asuka@jaxa.jp}

\author[0000-0002-6068-6337]{Takuya Miyazawa} \affiliation{Okinawa Institute of Science and Technology Graduate University, Kunigami-gun, Okinawa, Japan} \email{takuya.miyazawa@oist.jp}

\author[0000-0002-6054-3432]{Takashi Okajima} \affiliation{NASA Goddard Space Flight Center, Greenbelt, MD 20771, USA} \email{takashi.okajima@nasa.gov} 

\author[0000-0001-7011-7229]{Mark Pearce} \affiliation{KTH Royal Institute of Technology, Department of Physics, 106 91 Stockholm, Sweden} \affiliation{The Oskar Klein Centre for Cosmoparticle Physics, AlbaNova University Center, 106 91 Stockholm, Sweden} \email{pearce@kth.se}

\author[0000-0002-1452-4142]{Brian F. Rauch} \affiliation{Physics Department and the McDonnell Center for the Space Sciences, Washington University in St. Louis, 1 Brookings Drive, Saint Louis, MO 63130, USA} \email{brauch@wustl.edu}

\author[]{Kentaro Shirahama} \affiliation{Department of Earth and Space Science, The University of Osaka, 1-1 Machikaneyama-cho, Toyonaka, Osaka 560-0043, Japan} \email{shirahama@ess.sci.osaka-u.ac.jp}

\author[0000-0003-0710-8893]{Sean Spooner} \affiliation{Department of Physics and Astronomy and Space Science Center, University of New Hampshire, 8 College Road, Durham, NH 03824, USA} \email{Sean.Spooner@unh.edu}

\author[0000-0001-6314-5897]{Hiromitsu Takahashi} \affiliation{Graduate School of Advanced Science and Engineering, Hiroshima University,  1-3-1 Kagamiyama, Higashi-Hiroshima, Hiroshima 739-8526, Japan} \email{hirotaka@astro.hiroshima-u.ac.jp}

\author[0000-0002-7962-4136]{Yuusuke Uchida} \affiliation{Institute of Space and Astronautical Science, Japan Aerospace Exploration Agency, 3-1-1 Yoshino-dai, Chuo-ku, Sagamihara, Kanagawa 252-5210, Japan} \email{uchida.yuusuke99@jaxa.jp}

\author[0000-0003-1471-2693]{Varun*} \affiliation{KTH Royal Institute of Technology, Department of Physics, 106 91 Stockholm, Sweden} \affiliation{The Oskar Klein Centre for Cosmoparticle Physics, AlbaNova University Center, 106 91 Stockholm, Sweden} \email{vvarun@kth.se}

\begin{abstract}

We present results from hard X-ray observations of \cyg{} in the hard state, conducted during, and proximate to, the July 2024 balloon flight of \xlcp{}. Temporal and spectral observations from \nic, \nus, and \sw-\bat{} have been included, complementing \xlcp\ polarimetric observations in the $\sim$19--64 keV range. We perform joint \nic+\nus+\sw-\bat{} spectral analyses in the 1.0--195.0~keV range and reference these to the day-by-day polarization analysis. \xlcp{} previously reported (5.0$^{+2.7}_{-3.0}$)\% polarization degree (PD) and polarization angle (PA) aligned with the jet axis, when integrating over the four observing days. In this paper, we present day-by-day polarimetry results, which indicate variation in the polarization parameters. For July 12-13, the PD is ($16.8\pm4.9$)\%, at PA $\sim$25$^{\circ}$ shifted from the jet, while for the remaining days, the PD and PA are consistent with the integrated values. At the same time, we observed that the reflection fraction parameter ($\mathrm{R_{f}}$) of spectroscopic model increases by factor of two. These changes place specific requirements on the PD and flux of additional component responsible for them. Such requirements are more likely to be satisfied by geometric changes in corona near the BH than by scattering in the vicinity of the stellar companion.

\end{abstract}


\section{Introduction}
\label{sec:Introduction}

Black hole X-ray binaries (BHXRBs), where a stellar-mass black hole (BH) accretes from a companion star, show prominent emission variability on different timescales. Studies of these can provide useful diagnostics of BH accretion processes. When the companion star is an O or B-type supergiant, the variability of the high-energy emission may be driven by changes in the stellar-wind. Most BHXRB sources, like \gx, \xto{ }and \xtt, are transients that move dynamically in the hardness-intensity plane, transitioning between hard, soft and intermediate states on timescales of weeks to months  \citep{2010LNP...794...53B}. For some persistent sources like \lmo, \lmt{ }and \cyg, transitions are confined to a small region of the hardness-intensity plane \citep{2010LNP...794...53B}. The emission in the soft state is mainly thermal, from the accretion disk, whereas in the hard state, non-thermal emission from a hot coronal plasma is more prominent. The spectral properties in the intermediate state are mixture of the soft and hard states. The accretion-flow changes during the state transitions have not been identified. Furthermore, the shape, location, and physical properties of the corona responsible for the hard emission are still a matter of debate.   

\cyg\ is one of the brightest X-ray sources, located at a distance of ($2.22^{+0.18}_{-0.17}$) kpc, and it harbors one of the most massive black holes known among binary systems, with $M_\mathrm{BH} = (21.2\pm2.1) M_\mathrm{\odot}$~\citep{2021Sci...371.1046M}. This BH is in a binary orbit with an O-type supergiant companion  of mass ($40.6^{+7.7}_{-7.1}) M_\mathrm{\odot}$, with an orbital period of $\sim$5.6 days \citep{2015A&A...576A.117G}. Recent combined optical/UV studies have reported the black-hole mass to be lower however: in the range $12.7M_{\odot}-17.8 M_{\odot}$, depending on the binary inclination \citep{2025A&A...698A..37R}. Some high-energy observations have indicated that the BH in \cyg\ is nearly maximally rotating, i.e., having the dimensionless spin parameter $a_*$ close to unity~\citep{2011AAS...21841017G,2012MNRAS.424..217F,2014ApJ...780...78T}. Other works have found spin values in the range $0.80 - 0.99$ using different models, fitting conventions and assumptions about the source geometry \citep{2017PASJ...69...36K,2024ApJ...967L...9Z}.  The binary as a whole has an inclination $i=(27.0\pm0.8)^{\circ}$, as inferred from early optical observations \citep{2011ApJ...742...84O}. However, the hotter inner disk has been suggested to have a higher inclination, in the $35^{\circ}-45^{\circ}$ range \citep{2016ApJ...826...87W,2022Sci...378..650K,2024ApJ...967L...9Z}, possibly due to a disk warp or a misalignment relative to the outer disk. While the source does transition between soft and hard states, it does not show the typical q-shaped variation in the hardness intensity diagram \citep{2010LNP...794...53B}, instead only covering part of this cycle \citep{2020ApJ...896..101L}. The behaviour more closely resembles that of transient black-hole-binary sources than persistent sources, despite the source itself being persistent.

In the soft state, the \cyg\ spectrum is characterised by thermal emission from an optically thick and geometrically thin accretion disk, while the hard state is dominated by a non-thermal power-law component peaking at $\sim$100 keV, likely originating from Compton scattering in the hot corona. Long-term spectral and temporal variability in \cyg\ has been studied extensively using data from  \rxt\ and more recently from \nic.  The long-term \rxt\ data reveals that \cyg\ spends most of its time ($>$75\%) in the hard state, with brief transitions to the intermediate and soft states \citep{2013A&A...554A..88G}. These transitions usually take few weeks but can sometimes occur in just a few hours \citep{2011A&A...533A...8B}.  In terms of temporal variations, \cyg\ exhibits X-ray flares and transient dipping events. Flares have been observed during  \enquote{failed state transitions} \citep{2003A&A...407.1039P}, whereas dipping events are caused by interactions between X-rays from the BH and the stellar wind \citep{2015A&A...576A.117G,2022MNRAS.512.2671L}. A standard way to characterise the source timing variability is through the power-density spectrum (PDS), which can be decomposed into several distinct Lorentzian components. \cyg\ shows variable absorption at different orbital phases, being most prominent at the superior conjunction \citep{2015A&A...576A.117G}.  

There are questions regarding this source that are still unresolved, despite extensive temporal and spectral coverage. Polarimetric observations can shed light on these long-standing issues. Such observations are already challenging our understanding of BH binaries. In the hard state, \cyg\ exhibits a PD of $(4.01\pm0.20)\%$ at 2--8~keV \citep{2022Sci...378..650K} and $(5.0^{+2.7}_{-3.0})\%$ in the $\sim$19--64~keV range~\citep{2025ApJ...994...37A}. The PD is found to be energy dependent, increasing from ($3.6\pm0.4$)$\%$ to ($5.7\pm0.9$)$\%$ within the \ixpe\ energy band. The PD value also depends on the spectral state, decreasing to $(1.99\pm0.13)\%$ in the soft state, roughly half of its hard-state value \citep{2024ApJ...969L..30S}. Finally, the PD exhibits a linear relation with the hardness ratio (HR), suggesting a gradual evolution rather than abrupt changes between spectral states \citep{2025A&A...701A.115K}. 

The soft-state polarization parallel to the radio jet has been explained by a high BH spin of $a_*\ge$ 0.96 leading to a predominance of reflected emission \citep{2024ApJ...969L..30S}. Alternatively, an outflowing warm or hot ion-electron wind can explain the PD and PA \citep{2025ApJ...993...54K,2026arXiv260310870N}. The high PD in the hard state result from an outflowing corona \citep{2023ApJ...949L..10P} or from the reflection of a disk wind \citep{2024MNRAS.527.7047T}. The time-integrated PA is aligned with the radio jet axis both in soft X-rays, $(-20.7\pm1.4)^\circ$\footnote{Angles measured from celestial north toward northeast in a clockwise sense take negative values. } from \ixpe\ and hard X-rays, $(-28\pm17)^\circ$ from \xlcp{}. However, the PA in the \ixpe{} band show significant orbital change in the hard state, which might be related to changes in the amount of scattering at an intra-binary structure in the source \citep{2025A&A...701A.115K}. Beyond the \xlcp\ energy range, \pgp\ reported an upper limit of 8.6\% (90\% confidence level) in the 19-181 keV range \citep{2018NatAs...2..652C}. In soft gamma-rays, as observed by \intg, \cyg\ shows much higher polarization during the hard state, PD $=(75\pm32)\%$, and PA is shifted by $\sim$$90^\circ$ \citep{2015ApJ...807...17R}.

There are several radiation mechanisms in a binary BH system that can produce polarized emission. These include synchrotron emission from electrons in the magnetic field of the jet, Compton scattering of seed photons from the accretion disk in the corona, and scattering of disk radiation in the jet/wind. Synchrotron emission from the jet is dominant in optical/UV/IR and soft gamma-ray bands, and may also contribute to hard X-rays in some circumstances~\citep{2013MNRAS.434.2380M, 2014MNRAS.438.2083R}. Compton scattering from the corona dominates the hard X-ray emission. Most theoretical models predict that the non-thermal spectral component and polarized emission originate from the Comptonization of seed photons from the accretion disk or from synchrotron emission \citep{1985A&A...143..374S,1993A&A...275..337P, 2010ApJ...712..908S, 2022ApJ...934....4K}. The predicted PD of 1-2$\%$ for BH systems is challenged by high values from X-ray observations by both \ixpe{} and \xlcp. Some reconciliation with the observations can be made by assuming that the X-ray-emitting regions in these models are highly inclined. Other models assume scattering off mildly relativistic cold plasma outflows \citep{1987ApJ...322..650B, 2024MNRAS.528L.157D} or bremsstrahlung and synchrotron emission from the jet wall \citep{2024Ap&SS.369...68M}. 

\xlcp{} \citep{Abarr.2021} is a long-duration balloon-borne mission that carries a hard X-ray spectro-polarimeter sensitive in the $\sim$19-64~keV energy range. It observed \cyg\ and the Crab during its week-long 2024 flight, launched from Esrange, Kiruna, Sweden, and terminated near Kugluktuk, Nunavut, Canada. The first set of papers reported polarimetric results from the two sources integrated over the duration of the flight \citep{2025MNRAS.540L..34A,2025ApJ...994...37A, JacobPaper}. Here, we instead focuses on the day-by-day polarization and spectral behavior of \cyg{}, adding relevant observations from \nic, \nus, and \sw{}. 

Our paper is organized as follows: In Section~\ref{sec:Simultaneous observations}, we take inventory of contemporaneous observations of the source. Section~\ref{sec:Cyg monitoring} shows long-term monitoring data of \cyg{}, which allows us to determine the state and behavior of the source during the \xlcp{} flight. In Section~\ref{sec:Polarimetric analysis}, we present polarimetric results from \xlcp{} organized day-by-day, indicating a significant deviation during one observation.  Section~\ref{sec:Spectral analysis} shows how we combine observations in order to fit broadband spectra of \cyg{} during key parts of the \xlcp{} flight, allowing us to identify changes in the model parameters describing the source. We conclude with a discussion given in Section~\ref{sec:Discussion}.

\section{Simultaneous observations and data reduction}
\label{sec:Simultaneous observations}

\subsection{\xlcp}

\xlcp{} measures polarization using the angular anisotropy of Compton-scattered photons. Source photons are focused by an X-ray mirror \citep{TsunemiXrayMirror} onto a beryllium scattering element surrounded by pixelated CZT detectors. The polarimeter (scatterer + detectors) is housed inside an thick BGO anticoincidence shield to suppress measurement backgrounds \citep{IYER2023167975}. The mirror has a 12~m focal length, and sources are acquired and tracked using the Wallops ArcSecond Pointer (WASP) \citep{StuchlikWASP}. An imaging CZT detector located underneath the scatterer allows offsets in the focal point to be quantified.

\begin{figure}
    \centering
    \includegraphics[width=.475\textwidth]{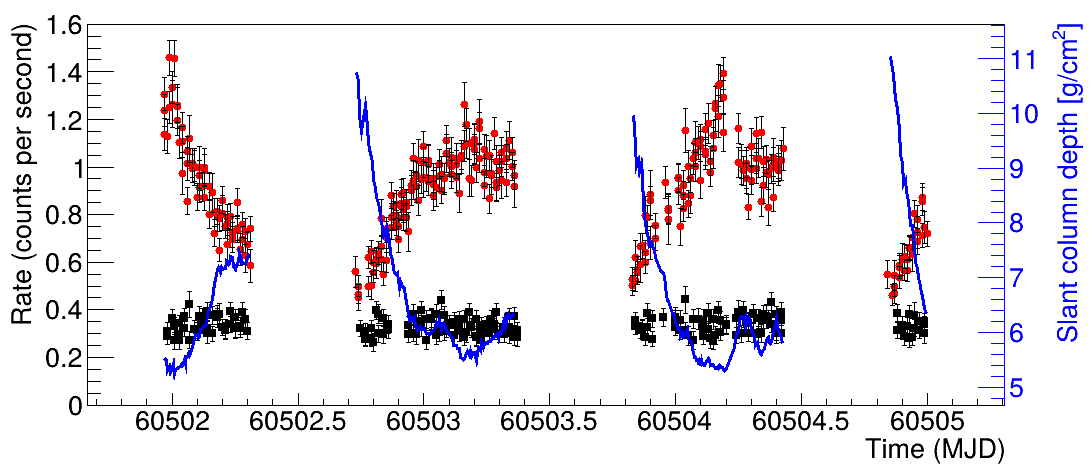}
    \caption{\xlcp\ observations of \cyg\ during the 2024 flight. Rates are shown for on-source (\enquote{signal}) events (red circles) and off-source (\enquote{background}) events (black squares). Four distinct periods are seen, where gaps between these correspond to periods where the Crab has been favoured for observation \citep{2025MNRAS.540L..34A, JacobPaper}. The signal rate can be seen to change as the source is rising and setting in the sky. Conversely, the background rate remains stable, despite considerable changes in the slant column depth (superimposed blue line and right-hand ordinate). During observations, diurnal variations in the altitude are between $\sim$36.5~km and $\sim$39.5~km.}
    \label{fig: Cyg visibility}
\end{figure}

During the 2024 flight of \xlcp, observations alternated between the Crab and Cyg X-1: since these sources are located at almost opposing right ascension, one will be rising as the other one is setting. Figure~\ref{fig: Cyg visibility} shows the four \cyg{} observations, conducted between July 10 and July 14. The first observation was truncated by the commissioning of the pointing system, and the last observation by the termination of science operations. Observations during July 11-12 and July 12-13 are mostly uninterrupted. Details of day-by-day observations\footnote{https://heasarc.gsfc.nasa.gov/FTP/xlcalibur/data/obs/2024/} for \cyg\ are given in Table~\ref{tab:obs}. For each day, the average offset of the mirror focal point was determined, allowing an offset correction to be applied to the determined scattering angles following the procedure detailed in~\cite{AoyagiSystematicEffects}.

\begin{table*}
    \caption{The observational data from \xlcp, \nic, and \nus\ used in this work.}
    \centering
    \hspace*{-2.3cm}
    \label{tab:obs}
    \begin{tabular}{|c|c|c|c|c|}
    \hline
     \textbf{Telescope}   & \textbf{Observation ID}        &  \textbf{Start time (UT)}        & \textbf{End time (UT)}     & \textbf{Exposure (sec)}   \\
    \hline
    \xlcp\      &  2024020002           & 2024-07-10 23:11:56       & 2024-07-11 07:29:50    & 14798         \\
                &  2024020003           & 2024-07-11 17:34:50       & 2024-07-12 08:37:45    & 27555         \\
                &  2024020004           & 2024-07-12 19:51:04        & 2024-07-13 10:16:40    & 20638         \\
                &  2024020005           & 2024-07-13 20:12:30       & 2024-07-14 00:15:51     & 6067          \\
    \hline
    \nic\       & 7706010110            & 2024-07-09 03:30:31        & 2024-07-09 23:40:26    & 4922      \\
                & 7706010111            & 2024-07-10 01:07:00        & 2024-07-10 22:53:46    & 4398      \\
                & 7706010112            & 2024-07-11 00:13:01        & 2024-07-11 23:39:26    & 4257      \\
                & 7706010113            & 2024-07-12 01:08:06        & 2024-07-12 21:08:06    & 817       \\
                & 7706010114            & 2024-07-13 00:04:35        & 2024-07-13 04:53:52    & 1871      \\
    \hline
    \nus\       & 91002320002           & 2024-07-10 15:00:30        & 2024-07-11 00:56:06   & 9419       \\
                & 91002320004           & 2024-07-12 15:03:32        & 2024-07-13 01:32:43   & 13720      \\
    \hline
    \end{tabular}
\end{table*}

\subsection{\nic}
\cyg\ was continuously monitored by \nic\ during the \xlcp\ campaign. The relevant \nic\ dataset consists of five observation IDs, one for each day of the \xlcp\ observations. Data reduction was done with \verb|NICERDAS 2025-09-05_V015|\footnote{https://heasarc.gsfc.nasa.gov/docs/software/lheasoft/help/nicer.html} and \verb|CALDB|, version 2024 February 06. All observations were processed with the \verb|nicerl2| task, which performs standard calibrations, screening and filtering and creates cleaned event lists. Relevant geomagnetic data was provided when running \verb|nicerl2| for the background-model creation\footnote{Data extraction was done as per the recommendation by the \nic\ team for observations done after May 22 2023, which are affected by an optical light-leak issue.}. Extraction of user products (light curves and spectral files) was done separately for \enquote{orbit night} and \enquote{orbit day} for the telescope. Data from the orbit night resulted in no good time intervals (GTIs) for \cyg. However, there were several GTIs for \cyg\ during orbit day, which allowed analysis using recommended selection cuts/criteria. Light curves were extracted in the 0.5-10.0 keV band with a time resolution of 10~s. These light curves have combined rates of 250-650~counts per second from all units. 

Due to the brightness of \cyg{}, the background spectral components are 3-5 times weaker than the source spectrum, hence a fixed background (instead of a parametrized background spectral model) was used for the spectral analysis. For the joint spectral analysis, we created GTIs for simultaneous \nic\ and \nus\ observations using the \verb|maketime| tool, and extracted further spectral products using the \verb|nicerl3-spect| tool. 

\subsection{\nus}
Two \nus\ observations of \cyg\ took place during the \xlcp\ flight. The first observation, July 10-11, coincides with an X-ray flare of the source. The second observation was from July 12, with a small portion extending into the next day. Details of these observations are given in Table \ref{tab:obs}. Processing of \nus\ observations was done using \verb|NuSTAR v2.1.5|, released on March 11 2025, and \verb|CALDB| from June 30 2025. Scripts {\verb|NUPIPILINE|} and \verb|NUPRODUCTS| were used to extract cleaned event files and user products \citep{2014AAS...22343805F}. We extracted all the events from a circular region of 180$^{\prime\prime}$ centered on \cyg\ for source spectrum files. A circular region of same size from a blank region in the image plane was used for the background spectral files. Spectral data were binned to have a minimum of 25 counts in each bin. Light curves were extracted in the 3--79~keV band.

\subsection{\sw-\bat}

Data products from the \bat\ all-sky monitor onboard \sw\ were analyzed with the \verb|BatAnalysis|\footnote{https://github.com/parsotat/BatAnalysis} package \citep{2023ApJ...953..155P}. This package allows downloading and processing of \bat\ survey data, consisting of events recorded when the \bat\ is not triggered by transients. The survey data can be used for monitoring of catalogued and uncatalogued hard X-ray sources in 14--195~keV energy range. We queried the HEASARC\footnote{https://heasarc.gsfc.nasa.gov/} database with the coordinates of \cyg\ during the dates of \xlcp\ flight. Furthermore, the \verb|SwiftBat| package\footnote{https://github.com/lanl/swiftbat$\_$python/} was used to select observations with detector-plane exposure area $>$1000~cm$^2$, corresponding to a partial coding fraction of $\sim$19$\%$ at the \cyg\ location. This \bat{}-specific criterion resulted in 47 observations suitable for our analysis. 

We cleaned regions for bright sources with signal-to-noise ratio (SNR) $\ge$6 and applied corresponding pattern maps. The data was rebinned into 8 energy bins for generating spectral files and response matrices. The resulting spectral data was then fitted with a simple power-law model in the 14--195 keV range. If the fitting was not good for particular observation ID or if \cyg\ was not detected at 3$\sigma$ above the background level, then a 5$\sigma$ upper limit of the flux was calculated \citep{2022ApJ...929..173L}. The photon index $\Gamma$ was fixed to 1.65 in this case, as expected for \cyg\ in the hard state \citep{2005MNRAS.363.1349G}. \verb|BatAnalysis| also calculates the source count rate, background count rate and SNR in each of the eight energy bins, which are then combined to find the net number of counts from \cyg\ in a given observation ID. 

\section{\cyg\ monitoring}
\label{sec:Cyg monitoring}
\begin{figure}
    \centering
    \includegraphics[width=0.98\linewidth]{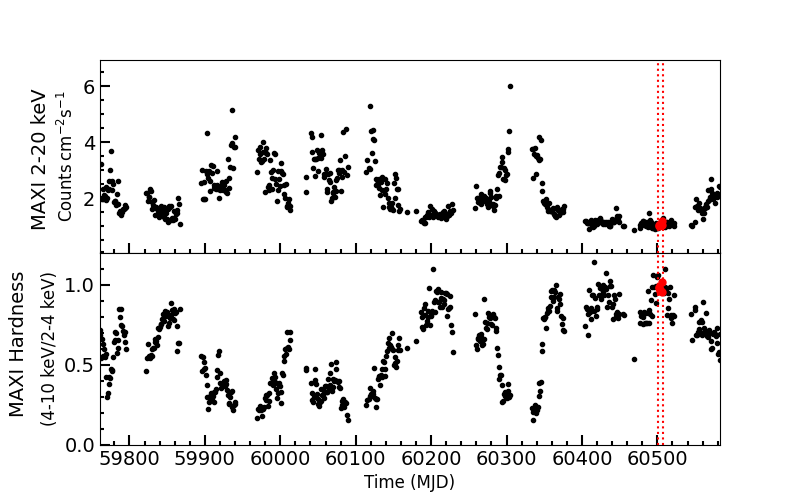}
    \caption{\cyg\ long-term monitoring from \maxi{} from July 1, 2022 to September 30, 2024 with light curve in the full 2--20 keV range (top) and hardness ratio between the 4--10~keV and 2--4~keV energy bands (bottom). Dotted red lines indicate the $\sim$week-long duration of the \xlcp\ observations.}
    \label{fig:ltlc}
\end{figure}

To characterize the long-term behavior of \cyg{}, we used \maxi\ \citep{2009PASJ...61..999M} data between July 1, 2022 and September 30, 2024 (MJD 59761-60583). Figure~\ref{fig:ltlc} shows the count-rate history in the 2--20~keV range, as well as the hardness ratio (HR) between 4--10 keV and 2--4 keV count-rates, over a two-year period. During the period covered, the source alternated between hard ($\mathrm{HR}>$0.5) and soft states ($\mathrm{HR}\le$0.5). The 2024 \xlcp\ campaign took place in a steady hard state, starting around MJD~60350. During observations, the hardness ratio was close to its peak value. 

\begin{figure}
    \centering
    \includegraphics[width=0.98\linewidth]{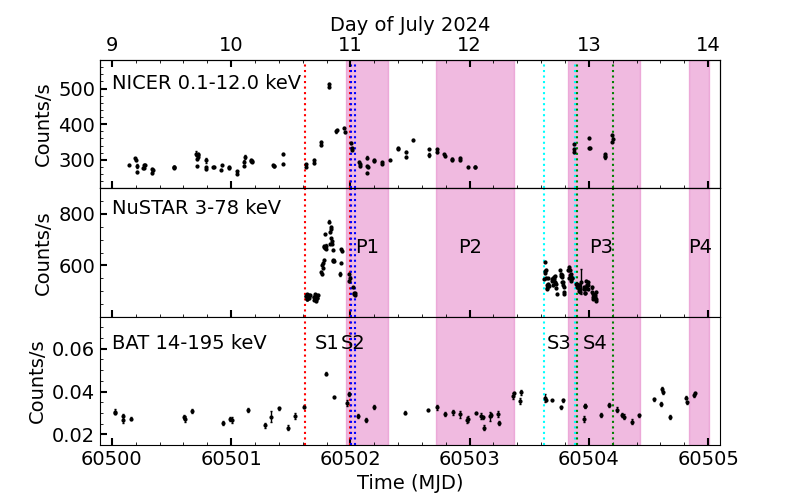}
    \caption{X-ray light curves of \cyg\ during July 9 to 14, 2024. Data from \nic\ (top) and \nus\ (middle) have a binning of 300~s. Each point in the \sw-\bat\ (bottom) light curve represents one observation ID from the survey data, which can have different exposures/durations. Vertical dotted lines in red (S1), blue (S2), cyan (S3), and green (S4) indicate four time segments that we use for our spectral (S) fitting. Shaded pink regions (P1-P4) show \xlcp\ polarimetric (P) observations.}
    \label{fig:lc}
\end{figure}

Figure~\ref{fig:lc} shows the \nic{}, \nus{}, and \sw{}-\bat{} light curves of \cyg{} during the \xlcp{} flight. The \nic{} count rate (0.1--12.0~keV) is steady around 300 counts/s on July 9 as well as in the first half of July 10. There is small flare late on July 10, lasting into the early hours of July 11, during which a peak count rate above 500~counts/s is observed. This is followed by a broader bump in the count rate on July 11. During July 12 and 13, the \nic\ light curve is covering \cyg{} only partially. The first \nus\ (3-78~keV) observation  covered the flaring activity on July 10-11, whereas the second observation shows a decrease, from 500-600~counts/s in the first half to below 500~counts/s in the second half. Finally, the \bat\ survey data (14-195~keV) gives uniform coverage of \cyg\ during the entire \xlcp\ campaign. X-ray flaring is also visible in the \bat\ light curve. Elevated count rates are seen during July 12 (where \nic\ data is missing), followed by a decay late on July 12 and early into July 13, matching the decrease seen in \nus{}. Variations in the \bat{} rate persist beyond the end of the second \nus{} observation.

\section{Polarimetric analysis}
\label{sec:Polarimetric analysis}
Using the procedure described in~\cite{2025MNRAS.540L..34A}, polarimetric results have been determined on a day-by-day basis, with results as presented in Table~\ref{tab: Day-by-day results}. Combining these observations yields the time-integrated result PD = (5.0$^{+2.7}_{-3.0})$\% at PA = ($-28 \pm 17)^{\circ}$ in the $\sim$19--64~keV range, as presented in~\cite{2025ApJ...994...37A}.

\begin{table*}
    \caption{Day-by-day polarimetric results from \xlcp{}. Intensity-normalized Stokes parameters Q/I and U/I take into account the specific modulation response for \cyg\ in the hard state \citep{AoyagiSystematicEffects} . PD and PA result from the marginalized posterior distributions of the Bayesian analysis \citep[see][and references therein]{2025MNRAS.540L..34A}. Uncertainties represent 1$\sigma$ Gaussian probability content. The minimum detectable polarization (MDP) \citep{WeisskopfMDP} is defined at a 99\% level.}
    \centering
    \hspace*{-3cm}
    \label{tab: Day-by-day results}
    \begin{tabular}{|c|c|c|c|c|c|c|c|}
    \hline
    \textbf{Obs. no.} & \textbf{Date} & \textbf{Q/I} & \textbf{U/I} & \textbf{PD (\%)} & \textbf{PA ($^{\circ}$)} & {\bf MDP (\%)}\\
    \hline
    1 & July 10-11 & 0.063 $\pm$ 0.053 & 0.006 $\pm$ 0.053 & 0.0$^{+7.4}_{-0.0}$ & 2.9 $\pm$ 33.7 & 16\\
    2 & July 11-12 & 0.080 $\pm$ 0.041 & -0.023 $\pm$ 0.041 & 7.0$^{+3.8}_{-4.8}$ & -8.1 $\pm$ 18.5 & 13\\
    3 & July 12-13 & -0.040 $\pm$ 0.048 & -0.170 $\pm$ 0.048 & 16.8$\pm$4.9 & -51.6 $\pm$ 8.4 & 15\\
    4 & July 13-14 & 0.063 $\pm$ 0.134 & 0.100 $\pm$ 0.134 & 0.0$^{+16.3}_{-0.0}$ & 28.9 $\pm$ 41.2 & 41\\
    \hline
    \end{tabular}
\end{table*}

\begin{figure}
    \centering
    \includegraphics[width=.475\textwidth]{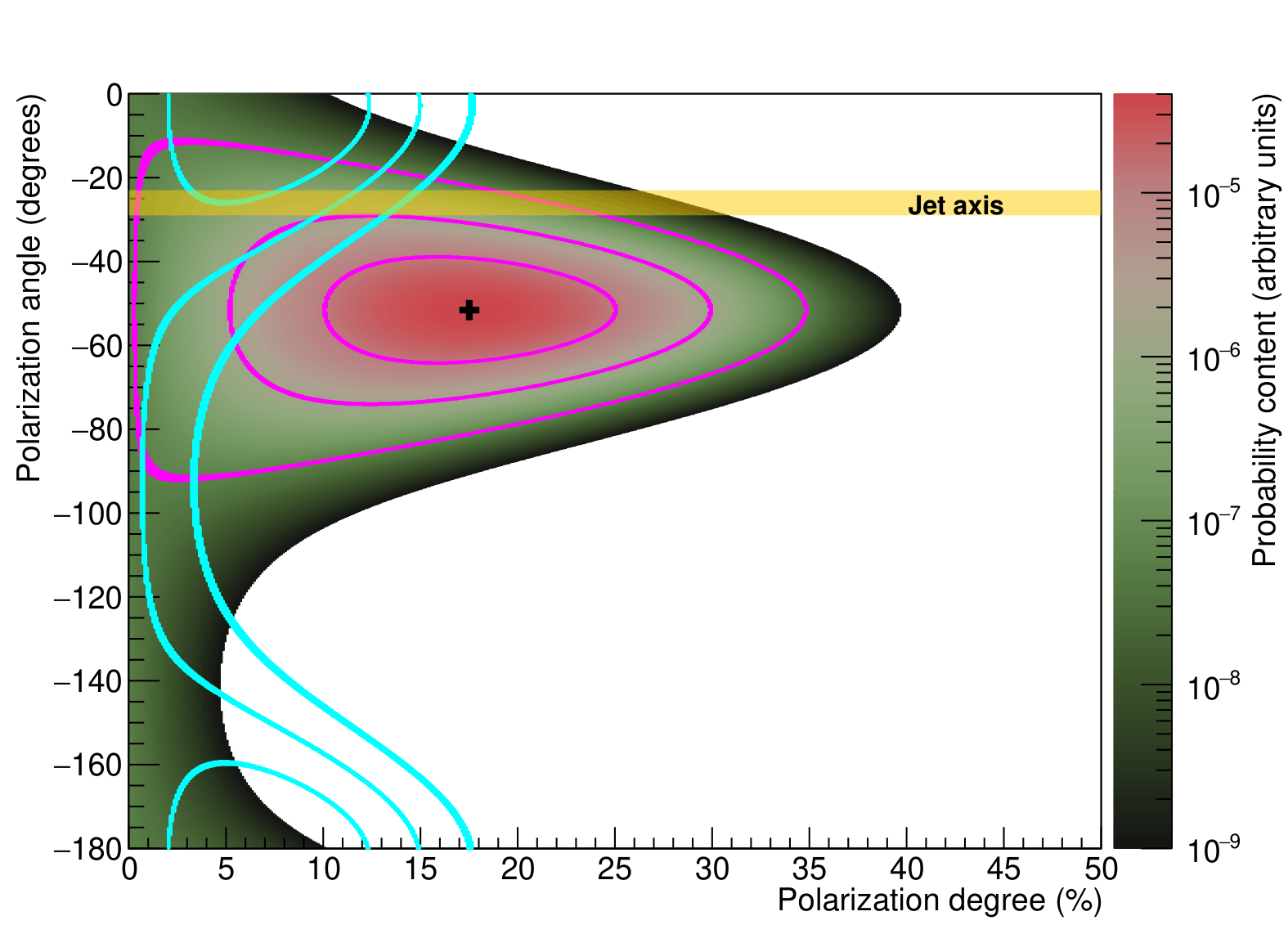}
    \caption{PD/PA comparison between observation 3 and the remaining datasets combined. The shaded region is the posterior density for observation 3 resulting from the Bayesian analysis \citep[see][and references therein]{2025MNRAS.540L..34A}, where the color scale on the right represents the probability content per bin/cell of the distribution. A black marker illustrates the \emph{maximum a posteriori} (MAP) result, corresponding to the bin/cell with the highest probability content: (PD, PA) = (17.5\%, $-51.6^{\circ}$). As expected \citep[see, e.g.,][]{XRayHandbook}, it corresponds to a higher PD than the marginalized value PD~=~16.8\%. Magenta lines indicate the 1, 2, and 3$\sigma$ contours (credibility regions) for observation 3, meaning 68.27\%, 95.45\%, and 99.73\% probability content, respectively. Cyan lines are the corresponding contours for the remaining observations combined. The semi-transparent yellow overlay indicates the jet direction from~\cite{2021Sci...371.1046M}. }
    \label{fig:Bayesian overlay}
\end{figure}

The PD = $(16.8\pm4.9)$\% in the July 12-13 observation (segment P3 in Figure 3.) is constrained at the $\sim$3.4$\sigma$ level. This individual value is $\sim$1.8$\sigma$ higher than the combined PD value from the remaining \xlcp{} observations: ($6.3^{+3.2}_{-3.6}$\%). The July 12-13 value exceeds the Minimum Detectable Polarization\footnote{Measurements below MDP can still provide meaningful constraints if a Bayesian analysis is used. See, e.g.~\cite{PolarizationConstraintsBelowMDP} for a discussion.} defined at 99\% confidence level \citep[MDP,][]{WeisskopfMDP}. The PA, which has a stronger constraint in this observation due to the high PD\footnote{The uncertainty in PA is inversely proportional to the PD, see, e.g.,~\cite{XRayHandbook}.}, is $(-51.4\pm8.4)^\circ$, a 24$^{\circ}$ offset  from the expected jet direction as determined using observations from the Very Long Baseline Array at 8.4~GHz \citep{2021Sci...371.1046M}. Figure~\ref{fig:Bayesian overlay} shows a comparison between observation 3 (segment P3) and the remaining datasets with using a Bayesian analysis.

\begin{figure}
    \centering
    \includegraphics[width=.475\textwidth]{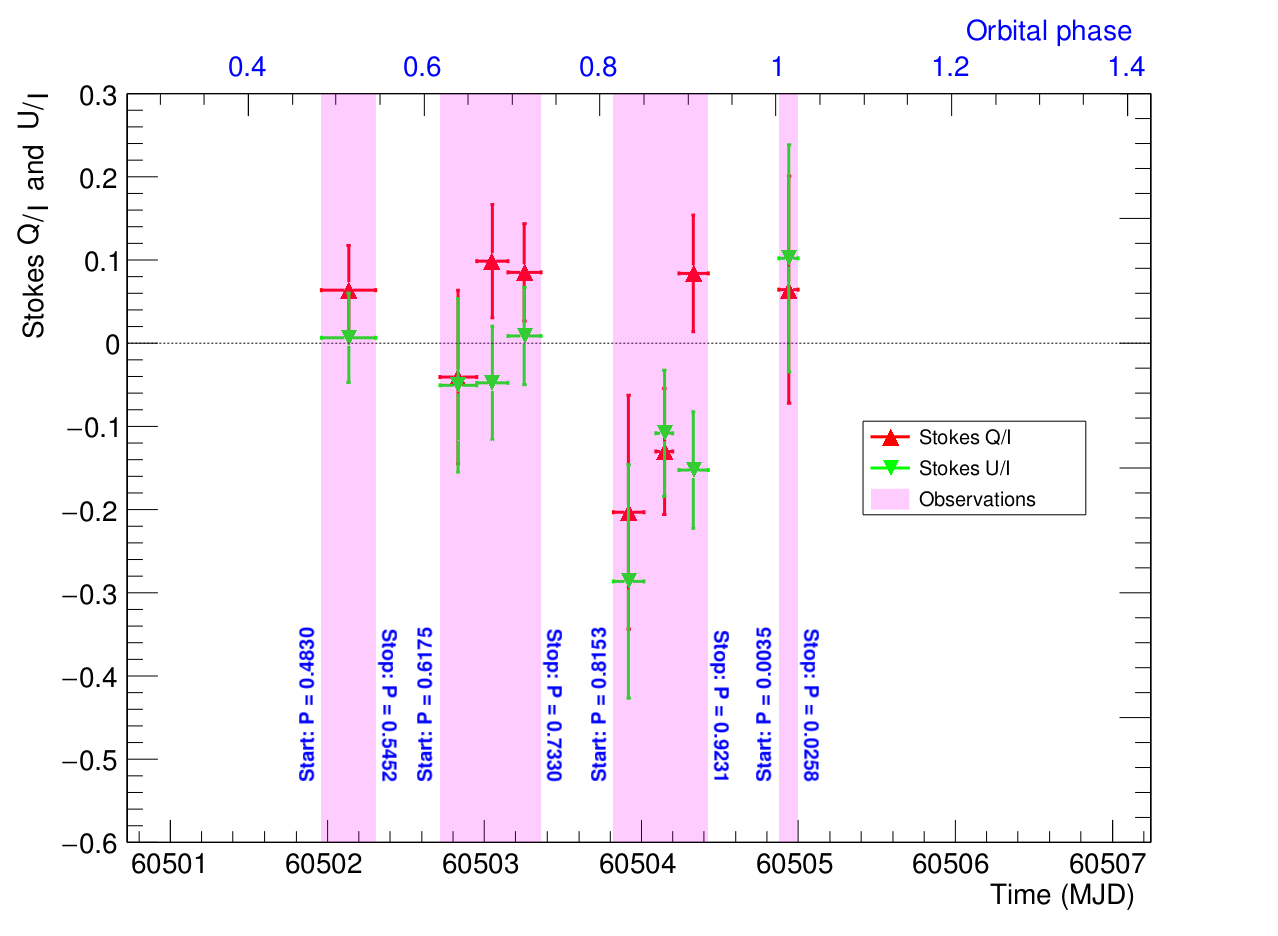}
    \caption{Intensity-normalized Stokes parameters Q/I (red triangles) and U/I (green inverted triangles) for the four \xlcp{} observations. A horizontal dotted line has been added at zero to guide the eye. Observations 2 and 3 have been sub-divided to provide better timing resolution. Horizontal error bars indicate the duration of each time interval. The top scale indicates the orbital phase~\citep{GiesOrbitalPeriod} of the binary system.}
    \label{fig:stokesvariability}
\end{figure}

Results in terms of intensity-normalized Stokes parameters are shown in Figure~\ref{fig:stokesvariability}. The Stokes parameters do not suffer from the bias due to the positive-definiteness of polarization measurements \citep{Quinn.2012, Maier_2014, mikhalev2018pitfalls} and therefore give a more robust way of identifying any variability. Data points corresponding to July 12-13 observation (segment P3) are clustered in different group than data points from other observations. Fitting the Stokes parameters with a constant yields a total $\chi^2=22$  for 14 degrees of freedom (dof), with a chance probability of $8\%$. Further, we checked the deviation of data points in segment P3 with respect to the remaining points. The combined local significance of variation of both parameters is $0.2\%$, indicating a strong significance. After correcting for trial factors, their global significance of $1.7\%$ also indicates significant variability. To confirm that the \xlcp{} measurement is not biased by changes in the polarimeter response or in the observational background, an extensive study has been performed, which is outlined in Appendix~\ref{app:Extended study}.

\section{Spectral analysis}
\label{sec:Spectral analysis}

The intriguing variation in polarization parameters could imply that physical changes are taking place in \cyg{} during the observation, in a way that changes the relative contributions of its emission components. Such changes would also affect the overall spectrum, allowing this idea to be investigated by studying broadband data from other instruments. We performed a joint spectral analysis for four different time segments motivated by variability in timing and polarization properties: \textbf{(S1)} during the flaring activity on July 10, \textbf{(S2)} during \xlcp\ observation on July 10-11, \textbf{(S3)} on July 12 before the start of \xlcp\ observation, and \textbf{(S4)} during the \xlcp\ observation on July 12-13. For the first two data sets, both \nic\ and \nus\ observed simultaneously. During the subsequent two segments, the overlap in coverage is only partial. Durations of these segments are indicated in Figure~\ref{fig:lc}. 

We used \verb|XSPEC| version 12.15.1 for fitting broadband spectra in the  1.0--195.0~keV range. Our full spectral model, in \verb|XSPEC| notation is \verb|MBPO| \verb|*| \verb|TBABS| \verb|*| \verb|(DISKBB| \verb|+| \verb|NTHCOMP| \verb|+| \verb|RELXILLCP| \verb|+| \verb|XILLVERCP)|. The \verb|MBPO| component accounts for the cross-calibration between different instruments and has the following functional form:

\begin{equation}
     \mathrm{M(E) = N(E/E_\mathrm{br})^{\Delta\Gamma}},
 \end{equation}

\noindent where N is a normalization based on the detection areas of the instruments at break energy, $\mathrm{E_\mathrm{br}}$, which is the point where the power-law spectrum changes from $\Delta\Gamma_1$ to $\Delta\Gamma_2$. For \nic, we fix the power-law indices of the \verb|MBPO| component to zero and its normalisation to unity. For \nus{} and \bat{}, we tie $\Delta\Gamma_2$=$\Delta\Gamma_1$ but keep the normalisation free. 

Continuum components \verb|DISKBB| and \verb|NTHCOMP| are available in \verb|XSPEC| standard models whereas the other additive components \verb|RELXILLCP| and \verb|XILLVERCP| are from the \verb|RELXILL| family of reflection table models \citep{2014ApJ...782...76G,2014MNRAS.444L.100D}. \verb|RELXILLCP| accounts for reflection of coronal X-rays off the inner disk and blurring by relativistic effects, whereas \verb|XILLVERCP| corresponds to X-ray reflection off outer parts of the accretion disk.

\begin{figure}
    \centering
    \includegraphics[width=0.98\linewidth, trim=6cm 1cm 10cm 3.1cm ]{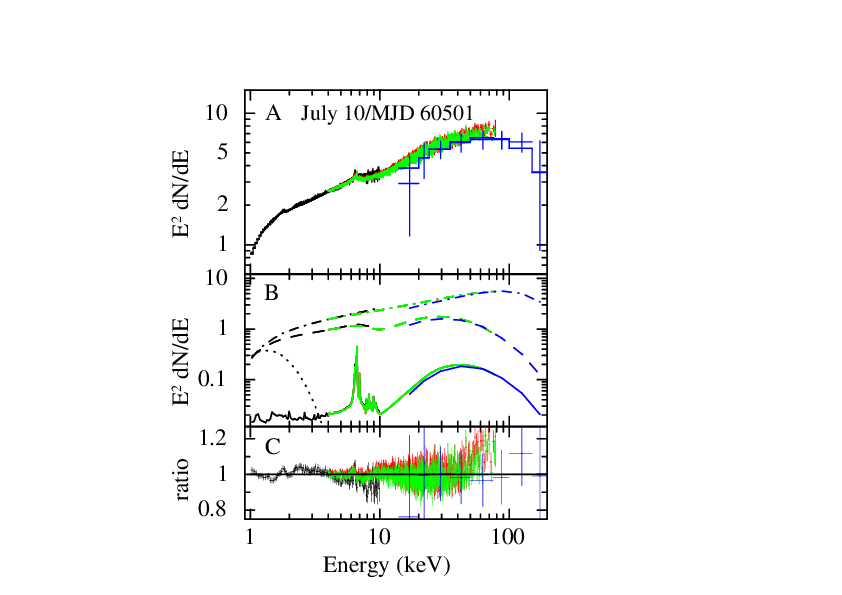}
    \caption{Spectral fitting of joint data for July 10. \textbf{(A)} \nic\ (black histogram), \nus\ (red and green for FPMA and FPMB), and \sw-\bat\ (blue) unfolded spectrum. The total model is shown by a solid line with colours matching the instrument data in the corresponding interval. \textbf{(B)} Additive components of the full spectral model: DISKBB, NTHCOMP, RELXILLCP and XILLVERCP are shown using dotted, dash-dotted, dashed, and solid lines, respectively. \textbf{(C)} residuals as the ratio between data and model.}
    \label{fig:sp}
\end{figure}

We tie the seed-photon temperature of \verb|NTHCOMP| ($\mathrm{kT_{bb}}$) to the accretion-disk temperature ($\mathrm{T_\mathrm{in}}$) of \verb|DISKBB|. Similarly, we fix the photon index ($\Gamma$) and electron temperature ($\mathrm{kT_{e}}$) of \verb|RELXILLCP| to the corresponding parameters of \verb|NTHCOMP|. The disk inclination is fixed to $38^\circ$ as obtained from the joint spectral analysis of data from \ixpe, \nic, \nus, and \intg\ \citep{2022Sci...378..650K}. Other disk-geometry parameters and emissivity-profile parameters were fixed at their default values in the model. We tie the normalization of \verb|RELXILLCP| to the \verb|NTHCOMP| component using a constant factor, determined such that the flux of the incident component of \verb|RELXILLCP| matches the corresponding \verb|NTHCOMP| flux in the 0.1--200.0~keV range. Parameter values from the initial fit are used for the flux calculation of \verb|NTHCOMP|, and this becomes the seed spectrum of \verb|RELXILLCP| for flux calculations. All parameters of \verb|XILLVERCP| were tied to corresponding \verb|RELXILLCP| parameters, except the ionization parameter ($\mathrm{log}\,\xi$) and its normalization. The resulting spectral fit for S1 is shown in Figure~\ref{fig:sp}.

We determine parameter errors as the $90\%$ confidence limits from Markov-Chain Monte-Carlo (MCMC) simulations \citep{20002dared}. These chains use the Goodman-Weare algorithm \citep{2010CAMCS...5...65G} with a total length of 220000 steps, spread over 110 walkers. From the first run of chains we determined that $\sim$55000 steps are needed for the values to stabilize. Subsequent runs used this as the burn-in length of chains before calculating errors. Resulting parameter values for the four different segments are given in Table \ref{tab:spec}.

\begin{table*}
    \centering
    \caption{ Best-fit parameter values for the joint spectral fits in the four time segments S1, S2, S3, and S4. Errors indicate the $90\%$ confidence intervals of parameters calculated from the MCMC chains.}
    \label{tab:spec}
    \begin{tabular}{|c|c|c|c|c|c|}
    \hline
     \textbf{Component} & \textbf{Parameter} & \textbf{Observation S1} & \textbf{Observation S2} & \textbf{Observation S3} & \textbf{Observation S4}             \\
    \hline
    \verb|TBABS|         &  $\mathrm{N_{H}}$             & $0.32_{-0.03}^{+0.04}$ & $0.28_{-0.04}^{+0.04}$ & $0.39_{-0.04}^{+0.06}$     & $0.34_{-0.03}^{+0.03}$ \\
    \hline
    \verb|DISKBB|        &  $\mathrm{T_{in}}$              & $0.38_{-0.04}^{+0.04}$ & $0.42_{-0.06}^{+0.08}$ & $0.19_{-0.03}^{+0.06}$     & $0.36_{-0.03}^{+0.04}$ \\
                  &  $\mathrm{Norm} (10^3)$       & $2.5_{-1.0}^{+3.4}$    & $1.1_{-0.5}^{+1.1}$    & $33_{-32}^{+122}$          & $3.0_{-1.3}^{+2.5}$ \\
    \hline
     \verb|NTHCOMP|      & $\Gamma$             & $1.65_{-0.02}^{+0.01}$  & $1.64_{-0.05}^{+0.04}$ & $1.58_{-0.02}^{+0.02}$    & $1.64_{-0.02}^{+0.02}$ \\
                  & $\mathrm{kT_{e}}$               & $228_{-119}^{+186}$     & $65_{-20}^{+56}$       & $107^{-49}_{+84}$         & $597^{-368}_{+360}$  \\
                  & Norm                 & $1.15_{-0.07}^{+0.06}$  & $1.0_{-0.2}^{+0.2}$    & $1.16_{-0.10}^{+0.06}$    & $1.12_{-0.05}^{+0.06}$ \\
    \hline
     \verb|RELXILLCP|   &log$\,\xi$               & $3.18_{-0.05}^{+0.18}$  & $3.2_{-0.3}^{+0.3}$    & $3.17_{-0.07}^{+0.20}$    & $3.17_{-0.05}^{+0.07}$\\
                 &$\mathrm{A_{Fe}}$               & $3.8_{-0.3}^{+0.4}$    & $4_{-1}^{+2}$          & $4.1_{-0.6}^{+0.5}$        & $4.0_{-0.4}^{+0.4}$ \\
                 &$\mathrm{R_\mathrm{f}}$                  & $0.16_{-0.03}^{+0.03}$ &  $0.11_{-0.03}^{+0.05}$ & $0.11_{-0.02}^{+0.02}$    & $0.25_{-0.03}^{+0.04}$ \\
    \hline
     \verb|XILLVERCP|   & log$\,\xi$              &  $2.0_{-0.4}^{+0.3}$   & $1.7_{-1.5}^{+1.5}$      & $1.4_{-1.2}^{+0.7}$      & $1.5_{-1.2}^{+0.7}$ \\
                 & Norm ($10^{-3}$)      & $5_{-1}^{+1}$          & $3_{-1}^{+1}$            & $3_{-1}^{+1}$            & $3_{-1}^{+1}$ \\
     \hline
    Calibration & $\mathrm{\Delta\Gamma_{\,Nustar}}$     & $-0.05_{-0.02}^{+0.02}$  & $-0.063_{-0.060}^{+0.047}$ & $-0.12_{-0.02}^{+0.03}$ & $-0.05_{-0.02}^{+0.02}$ \\
                & $\mathrm{Norm_{\,Nustar}}$       & $1.002_{-0.006}^{+0.005}$ & $0.94_{-0.01}^{+0.01}$ & $0.962_{-0.007}^{+0.007}$  & $0.972_{-0.005}^{+0.005}$  \\
                & $\mathrm{\Delta\Gamma_\mathrm{\,BAT}}$        & $-0.11_{-0.15}^{+0.14}$  & $0.02_{-0.26}^{+0.29}$   & $-0.13_{-0.31}^{+0.24}$   & $-0.1_{-0.1}^{+0.1}$ \\
                & $\mathrm{Norm_{\,BAT}}$          & $1.2_{-0.4}^{+0.6}$      &  $0.6_{-0.3}^{+0.5}$     &  $0.8_{-0.4}^{+0.8}$      & $0.9_{-0.2}^{+0.3}$ \\
    \hline
    \end{tabular}
    
\end{table*}

Spectral parameters obtained from these fits confirm that \cyg\ was in the hard spectral state during the \xlcp\ campaign with spectral index $\sim$1.6, the same hard-state value as obtained in previous works \citep{2015ApJ...808....9P, 2017MNRAS.472.4220B, 2022Sci...378..650K}. During the first two segments, the parameter values for all spectral components agree within their uncertainties. The inner disk temperature $\mathrm{T_{in}\sim}$0.4 indicates that this component contributes only in the soft X-ray part of the spectrum relevant for \nic{}. Conversely, the non-thermal component, with spectral index $\Gamma=1.64$ and electron temperature $\mathrm{kT_{e}>}$70 keV, dominates in the hard X-ray band covered by \nus\ and \bat{}. The relativistic reflection component shows that a highly ionized inner disk (log$\,\xi\sim$3.2) also contributes strongly to the high-energy emission, with reflection fraction $\mathrm{R_{f}}\sim$0.13. Finally, the non-relativistic reflection component is less ionized, with log$\,\xi\sim$1.7, and contributes less to the spectra. 

During segment S3, the disk temperature decreases with an increase in normalization (poorly constrained). At the same time, the spectral index ($\Gamma$) also changes to 1.58 indicating that the spectrum is harder in this segment. Both reflection components show similar values as for segments S1 and S2. In the last spectral segment, S4, which coincides with the \xlcp\ observation with the polarization excursion, more spectral parameters show change. First, we note that $\Gamma$ increases from its value in the previous segment back to a value consistent with the first two segments. The electron temperature parameter ($\mathrm{kT_{e}}$) is higher in this segment than all previous segments, but is not well constrained. Finally, the relativistic reflection fraction increased from $\mathrm{R_f} = 0.11\pm0.02$ (S3) to $\mathrm{R_f}=0.25^{+0.04}_{-0.03}$ (S4), while other parameters remain compatible within their uncertainties. To ensure that the $\mathrm{R_f}$ change is not an artifact from the poorly constrained $\mathrm{kT_e}$ parameter, we fixed this parameter to the value generally seen the hard state \citep{2017MNRAS.472.4220B, 2022Sci...378..650K}. The value thus obtained is $\mathrm{R_f}$  ($0.22^{+0.03}_{-0.03}$), which is still more than $3\sigma$ higher from the value for S3.

\section{Discussion}
\label{sec:Discussion}

Polarization estimates in BH sources go back to seminal work of \cite{1985A&A...143..374S}, a simple geometrically flat and optically thin Compton cloud, they showed that the high energy polarization depends only on the optical thickness of the Comptonizing medium and its inclination ($\mathrm{\mu = \cos{i}}$) to the disk normal. This framework allows for the calculation of the electron temperature ($\mathrm{k_{B}T_e}$) of the source using polarization and X-ray spectral measurements \citep{2025A&A...704A..87T}. The polarization of several stellar BHs such as \cyg{} are well described in this model and their derived $\mathrm{k_{B}T_{e}}$ follow the expected bimodal behavior in the hard and soft states. \xlcp{} results of \cyg{} (except P3) are consistent with the time-integrated value ($\sim5\%$) which corresponds to the lowest optical depth $\mathrm{\tau =1}$ for \cyg{} at $\mathrm{\mu} =0.9$, resulting in a typical hard state electron temperature ($\mathrm{k_{B}T_{e} = 70~keV}$). The PD during segment P3 is $16.8\%$ and cannot be obtained in this framework for low-inclination sources like \cyg{}. Indeed, the highest value $(16-18)\%$ can only be achieved for a very high inclination ($\mathrm{\mu = 0.1-0.3}$).

The more recent theoretical model \verb|kerrC| accounts for general-relativistic propagation effects through curved spacetime and the reflection off the accretion disk around the BH \citep{2022ApJ...934....4K}. This model predicts an energy dependence of PA in BH sources.
The PD from \xlcp{} \cyg{} observations, at around 5\%, was interpreted in \cite{2025ApJ...994...37A} in terms of the wedge-shaped corona geometry. The best agreement to both \xlcp{} and \ixpe{} PDs was obtained for an observer inclination angle of $i=55^{\circ}$ relative to the black-hole spin axis. Such a high inclination indicates a mismatch between the geometry of the binary system and that of the high-energy-emitting inner-disk region — suggesting the presence of a warped structure. While the PD generally rises with increasing $i$ in this theoretical framework, values greater than $10\%$ cannot be achieved. Furthermore, the PA is always aligned with the projected accretion-disk axis for the wedge-shaped configuration of the corona in this model. Finally, the fact that high polarization is obtained only for one out of the four observations indicates variability on the day-scale. Such variability is currently not supported in this model, which cover a static case.

\begin{figure}
    \centering
    \includegraphics[width=0.475\textwidth]{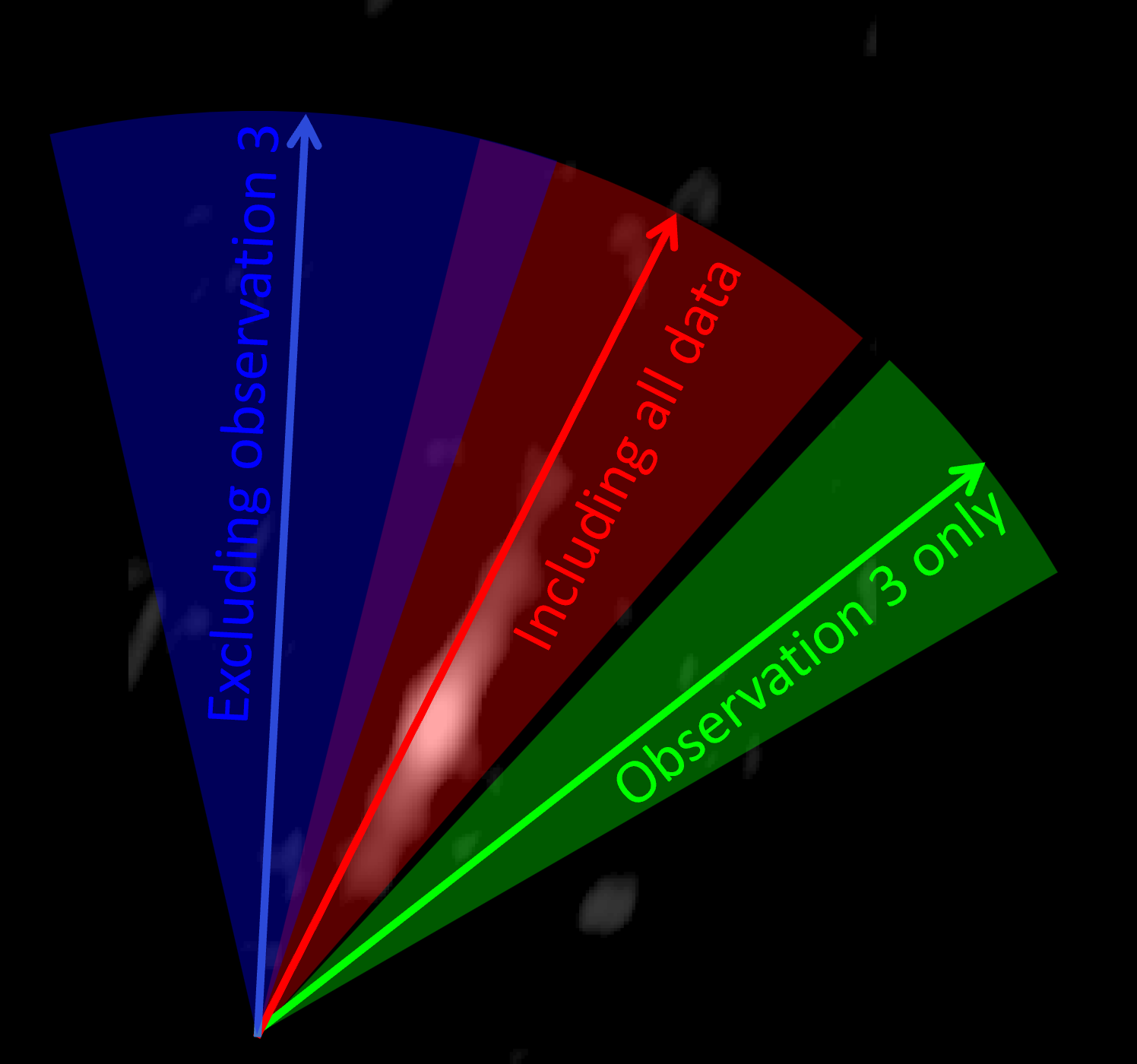}
    \caption{Polar diagram for \cyg{} polarization from \xlcp{} data overlaid on the radio jet axis. Arrows indicate measured PA values, with coloured bands corresponding to their 1$\sigma$ uncertainties. The red band shows the PA obtained from all observations, aligning with the jet axis as reported in \citep{2025ApJ...994...37A}. Green and blue bands are for the PA of \xlcp{} observation 3 only, and for all data excluding observation 3, respectively. The green and blue bands are thus statistically independent components of the dataset yielding the red band.}
    \label{fig:PA and jet axis}
\end{figure}

Although the polarization excursion in our observation is notably large, \ixpe{} results have already demonstrated that the polarization properties of \cyg{} are not static. \cite{2025A&A...701A.115K} found that its polarization changes with hardness ratio, spectral state, and orbital phase. Within the hard spectral state, PD is found to change by $\sim$$2\%$ around the binary orbit. The observed PA also varies, about half a cycle out of phase with respect to PD, and by $\lesssim$10$^{\circ}$. This behaviour is interpreted as a polarization vector moving on a circle in the intensity-normalised Stokes Q/I and U/I parameter space. A systematic search or confirmation of such an orbital variability is difficult here owing to partial coverage of the orbit ($\sim$70$\%$), gaps in data when observing the Crab system, irregular durations, and large errors when further sub-dividing individual \xlcp{} measurements. However,  PD = ($16.8\pm4.9$)$\%$ during segment P3, is more than 3 times as high as previously reported values for \cyg\ with \xlcp{} observations. Also, during the excursion, the PA shifts from the jet axis by more than one standard deviation. The PA of data excluding P3 is compatible with the direction of the radio jet.

Spectral variability in \cyg{} accompanied with polarization changes indicate physical changes in the X-ray-emitting region. \cyg\ is both bright and dynamic in nature, with variability in total flux as well as in its individual spectral components. Spectral studies of the reflection feature have evolved with the development of spectral models and the availability of the broadband spectral data. Relativistic reflection was modeled with \verb|REFLIONX|, using broadband \suz\ observations taken during the hard state in April-December 2009. The relativistic disk-reflection flux was found to vary between (0.44-4.19)$\times10^{-8}$ erg $\mathrm{cm^{-2}}$~$\mathrm{s^{-1}}$ \citep{2012ApJ...757...11M}. More recent works use the \verb|RELXILL| models with the disk-reflection component parametrized using the reflection fraction $\mathrm{R_f}$, defined as the ratio of the reflection component flux to the continuum component flux in the 20--40 keV band. \nus\ observations in 2012 and 2014 found that $\mathrm{R_f}$ increased by a factor of 3 between hard and soft spectral states \citep{2015ApJ...808....9P}. Ruling out the truncation of the disk from their spectral modelling, they suggested that such a change could be explained by an outflowing material or increased Comptonization of reflection emission by the corona. The inner disk parameter was not constrained in our spectral fitting and was fixed at the default value for all the spectral segments. 

In our spectral modelling, the \cyg{} emission comes from four different components. Out of these, Compton emission and relativistic-reflection emission dominate in terms of the total flux. The disk emission mainly contributes in low energies ($<$5 keV), and the non-relativistic reflection emission is three orders of magnitude lower than the Compton emission. Hence, their contribution to the total polarization can be neglected, and net polarization from the source can be given as:  

\begin{equation}
    \mathrm{PD(E)} = \frac{\mathrm{F_{Comp}(E)}\times\mathrm{PD_{Comp}(E)} + \mathrm{F_{refl}(E)}\times\mathrm{PD_{refl}(E)}}{\mathrm{F_{total}(E)}},
\end{equation}

\noindent where F indicates the flux and PD the polarization degree of the Comp(ton), refl(ection) and total components, respectively. The $\mathrm{R_f}$ of 0.25 in segment P3 and Compton-emission polarization of 6.3\% (from the integrated result of \xlcp{} excluding P3) would require the reflected emission to have PD $\sim$50$\%$ to yield the net PD of $16.8\%$ as observed with \xlcp{} during P3. In the following, we discuss two hypotheses to explain the observed behavior.

The first hypothesis involves changes close to the BH. The increase in $\mathrm{R_f}$ is most naturally interpreted as a change in coronal plasma and/or magnetic geometry. The Stellar corona are dynamic, with time-varying magnetic-field morphologies. Intuitively, one expects that the corona above the accretion disks should be similar in many respects. Changing the direction of the skeletal magnetic field that carries the coronal plasma will naturally increase the PD when an observer is viewing this cloud at a higher obliquity to the field direction.  At the same time, a dynamic field could move the corona relative to the disk and alter the fractional contribution of the polarized reflection component: higher PDs might suggest a corona closer to the disk. Such a picture motivates deep general-relativistic MHD studies of the expectations for the dynamic field morphology and polarization changes near corona/disk interfaces in black-hole systems. Explaining changes on timescales of a few days serves as a constraint on such models.

\begin{figure}
    \centering
    \includegraphics[width=0.475\textwidth]{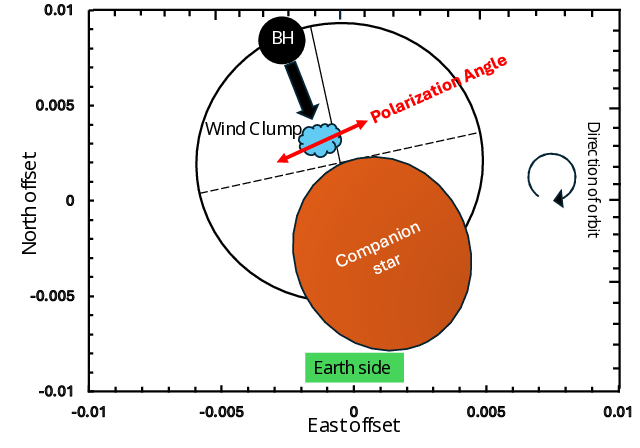}
    \caption{Schematic diagram of the relative positions of the different components of \cyg{} at phase $\sim$0.85, i.e., during \xlcp{} observation 3. Phase zero corresponds to the companion star being at the inferior conjunction. The orbit of the BH is an ellipse with a clockwise rotation. Major and minor axes are shown with solid black and dashed lines, respectively. Scattering off a clump along the line between the BH and the companion star can produce polarization with the direction as indicated by the red arrow.}
    \label{fig:wre}
\end{figure}

The second hypothesis involves scattering from the companion star and its stellar wind. The total effect of the scattering of the companion star and its homogeneous stellar wind induces a modulation of polarization that alternately reinforces and counteracts the source polarization throughout the orbit, as studied by \cite{2026arXiv260604159V} by combining a general-relativistic description of the polarized X-ray source emission (\verb|kerrC|) with a focused stellar wind model for the binary medium. Their results of PD modulation amplitudes are small in the \ixpe{} band, with an increase from $0.25\%$ at 2--4~keV to $1.24\%$ at 6--8~keV. Hence, the large polarization change seen in \xlcp{} is unlikely to be caused by these elements, despite the higher energy of these measurements. However, their simulations do not include the clumping and huge ionization range of the wind \citep{2005ApJ...620..398M,2016A&A...590A.114M}. These wind clumps are matter with low temperature and high density, surrounded by photo-ionized gas. Such a clumpy wind introduces variable absorption over the binary orbit of \cyg{}, with increased scattering in the phase range 0.8-1.3, near superior conjunction \citep{2015A&A...576A.117G}. The \xlcp{} observation with the excursion (P3) corresponds to an orbital phase of $\sim$0.85. \cyg{} is viewed at an inclination of $27^{\circ}$, with the angular-momentum vector pointing away from the observer as inferred from radio interferometry measurements \citep{2011ApJ...742...83R,2021Sci...371.1046M}. Considering the inferior conjunction of the companion star as the reference phase ($\phi=0$), the BH (companion) is away from (towards) the observer at a orbital phase $\phi\sim0.85$, as shown in Figure \ref{fig:wre}.
In this orientation, photons originally directed toward the companion star could be scattered towards the observer, leading to an enhancement of PD and  a change in PA away from the jet direction. However, an increase in PD requires that the scattered component also has an additional $10\%$ flux from the scattered component during P3 ($\mathrm{R_f =0.25}$ compared to 0.10-0.15 in other segments). A simple calculation considering Thompson scattering in a wind clump, size $\mathrm{r_c=10^{10}~cm}$ having electron density, $\mathrm{n_e=10^{10}~cm^{-3}}$, located in the vicinity of the stellar companion (distance from BH, $\mathrm{d_c=10^{12}~cm}$) gives the relative flux of $\sim$$10^{-10}$, many order of magnitude less than what would be required to produce this change in polarization. Hence, we conclude that this scenario is insufficient to cause the transient change.

The high polarization degree ($>$10$\%$) observed here increases the tension between observations and predictions based on current static theoretical models. Even among the sources observed by \ixpe{} since 2021, some of the polarization measurements are puzzling. One such case is the BH binary \bht{}, where PD increases to $10\%$ during the soft state. This cannot be easily reconciled with standard models despite the high inclination of the source \citep{2024ApJ...964...77R}. Similarly, some low-mass X-ray binaries with weakly magnetized neutron stars (WMNSs) like \bhi{} \citep{2025A&A...697A..83A}, \gxo{} \citep{2024A&A...691A.253L}, \ax{} \citep{2025arXiv251213182M} show unexpectedly high polarization. These findings clearly demonstrate that theoretical models must incorporate the intrinsically dynamic nature of binary astrophysical sources, including sudden variations in their accretion geometry. The findings also emphasize the crucial role of simultaneous multi-wavelength observations for characterizing and understanding these sources.

\section*{Acknowledgements}
{\it \xlcp} is a joint mission supported by NASA, JAXA, and the Swedish National Space Agency {\it Rymdstyrelsen}. We acknowledge NASA support under grant 80NSSC24K0205.
KTH authors are supported by the Swedish National Space Agency (2022-00178 and 2024-00248). MP also acknowledges funding from the Swedish Research Council {\it Vetenskapsrådet} (2021-05128 and 2025-04648). The Washington University in St. Louis group acknowledges additional NASA support through the grants 80NSSC20K0329, 80NSSC21K1817, 80NSSC22K1291, 80NSSC22K1883, 80NSSC23K1041, and 80NSSC24K1178, as well as funding from the McDonnell Center for the Space Sciences at Washington University in St. Louis. 
The University of New Hampshire group acknowledges additional NASA support through the grants 80NSSC24K0636, 80NSSC24K1762, and 80NSSC25K8001. The Japanese Society for the Promotion of Science (JSPS) has supported this work through KAKENHI Grant Numbers 19H01908, 19H05609, 20H00175 (HM), 20H00178 (HM), 21K13946 (YU), 22H01277 (YM), 23H00117, and 23H00128 (HM).




\appendix

\section{Systematic studies of \xlcp{} results}
\label{app:Extended study}
An extensive investigation has been conducted to verify that the excursion in polarization parameters seen during \xlcp{} observation 3 (July 12-13) is not caused by changes in the measurement background or by instrument effects. Studies include checking for changes in the rate of background particles registered by the anticoincidence shield \citep{IYER2023167975}, spurious noise in the detector, cross-contamination between signal and background (on-source/off-source) observations and validity of exposure normalizations. 

\begin{figure*}
    \centering
    \includegraphics[width=0.475\textwidth]{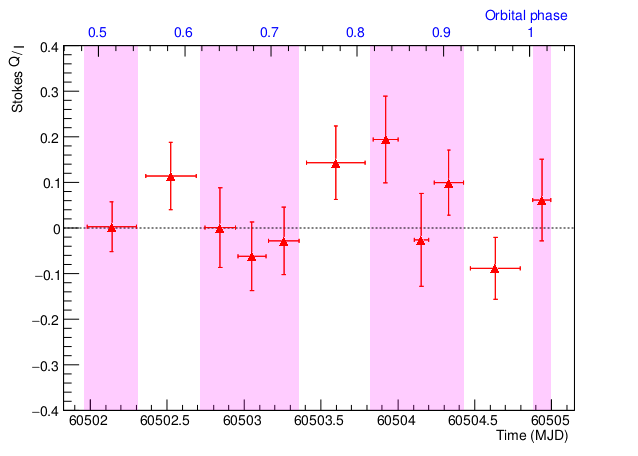}
    \includegraphics[width=0.475\textwidth]{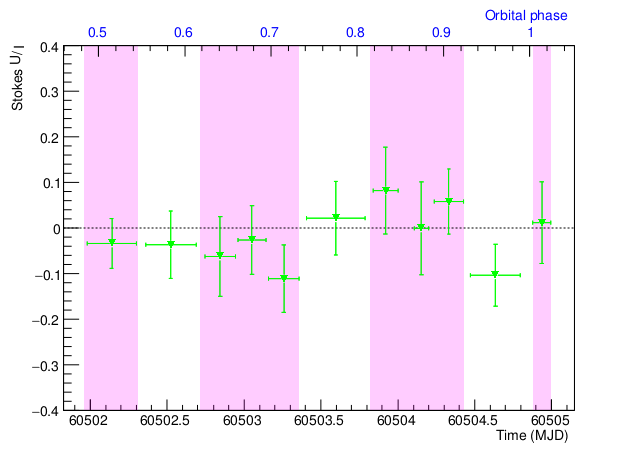}
    \caption{Intensity-normalized Stokes parameters Q/I (left) and U/I (right) from the \xlcp{} background observations. The top scale shows the orbital phase of \cyg{} \citep{GiesOrbitalPeriod}. Horizontal error bars indicate the duration of each time interval. Dashed lines have been added at Q~$=$~0 and U~$=$~0 to guide the eye. The four shaded bands indicate the \cyg{} observations, and data points between these are from the Crab \citep{2025MNRAS.540L..34A, JacobPaper}. To allow direct comparison, results have been calculated without using any offset correction \citep{AoyagiSystematicEffects}. The high signal-to-background ratio (see Figure~\ref{fig: Cyg visibility}) further reduces the impact of any variations in the background. }
    \label{fig:Background Stokes parameters}
\end{figure*}

As was shown in Figure~\ref{fig: Cyg visibility}, the background rate remained stable throughout the four observations, excluding a change caused by varying cosmic-ray flux or altitude. It also demonstrates that pixel noise, which is filtered in the analysis, is not altering measurements during the July 12-13 observation. For each pixel, the cumulative number of events has been inspected as function of time, to ensure that there are no sudden increases which could indicate pixels becoming noisy even in brief intervals. The intensity-normalized Stokes parameters of the background observations furthermore demonstrate that the nature of the background subtracted for this observation is not phenomenologically different from the other background sets. This is shown in Figure~\ref{fig:Background Stokes parameters}, where background results from the Crab observations \citep{2025MNRAS.540L..34A, JacobPaper} have also been included. 

A conventional modulation-curve approach has also been tested, where the angular distribution of Compton-scattering angles is fitted with a sinusoidal function with 180$^{\circ}$ and 360$^{\circ}$ components. Studying the scattering-angle distribution and the fitted components allows the measurement to be evaluated for behavior that calculated Stokes parameters Q and U are agnostic to, like the phase and amplitude of the 360$^{\circ}$ component.
As described in~\cite{AoyagiSystematicEffects}, the 180$^{\circ}$ component arises mainly from the polarization of the incident flux, while 360$^{\circ}$ component is caused by an offset of the focused flux relative to the optical axis of the X-ray detector. Fitting parameters corresponding the data in Figure~\ref{fig:stokesvariability} are shown in Table~\ref{tab:Fitting parameters - signal}. Also this method indicates an increase in PD during the July 12-13 observation, while the amplitude of the 360$^{\circ}$ component remains low, demonstrating the justification of the offset correction. Finally, fitting parameters for the background-only data of Figure~\ref{fig:Background Stokes parameters} are given in Table~\ref{tab:Fitting parameters - background}.

\begin{table*}
    \centering
    \caption{Fitting parameters for the scattering-angle distributions (\enquote{modulation curves}) of the \xlcp{} observations matching Figure~\ref{fig:stokesvariability}. Parameters $A$ and $\phi$ are the amplitude and phase of three fitted harmonics, corresponding to the 180$^{\circ}$, 360$^{\circ}$ and 120$^{\circ}$ components described in~\protect\cite{AoyagiSystematicEffects}. By definition, amplitudes are positive definite. Furthermore, phase values become poorly defined when amplitudes are low. The uncertainties reported here are fitting errors only and should not be taken as true measurement errors: they are included here to indicate the phenomenological similarity between the different datasets. The limitations of the modulation-fitting approach underline the benefit of a Bayesian analysis for determining the polarization parameters PD and PA and their associated errors (Bayesian credible regions). Even with the fitting parameters, the increased \enquote{PD} in the observation on July 12-13 becomes apparent from amplitude of the 180$^{\circ}$ component. The amplitude of the 360$^{\circ}$ component is higher in the result for July 10-11, as the \xlcp{} pointing system was still being fine-tuned during the initial part of the observation. Consequences of this fine-tuning are negated by applying an offset correction to the reconstructed scattering events, as explained in~\protect\cite{AoyagiSystematicEffects}.}
    \label{tab:Fitting parameters - signal}
    \begin{tabular}{|c|c|c|c|c|c|c|c|}
    \hline
    \textbf{Obs. no.} & \textbf{Date} & \textbf{$A_{180}$} & \textbf{$\phi_{180} (^{\circ})$} & \textbf{$A_{360}$} & \textbf{$\phi_{360} (^{\circ})$} & \textbf{$A_{120}$} & \textbf{$\phi_{120} (^{\circ})$} \\
    \hline
    1 & July 10-11 & 0.028 $\pm$ 0.023 & 87.5 $\pm$ 22.8 & 0.070 $\pm$ 0.023 & 129.3 $\pm$ 18.4 & 0.006 $\pm$ 0.023 & 105.2 $\pm$ 105.3 \\
    2\_1 & July 11-12 Part 1 & 0.035 $\pm$ 0.042 & 126.6 $\pm$ 33.5 & 0.086 $\pm$ 0.042 & 62.9 $\pm$ 27.8 & 0.052 $\pm$ 0.042 & 128.9 $\pm$ 81.8 \\
    2\_2 & July 11-12 Part 2 & 0.053 $\pm$ 0.029 & 101.0 $\pm$ 15.6 & 0.020 $\pm$ 0.029 & 180.1 $\pm$ 78.8 & 0.023 $\pm$ 0.029 & 8.9 $\pm$ 21.5 \\
    2\_3 & July 11-12 Part 3 & 0.038 $\pm$ 0.025 & 88.0 $\pm$ 18.7 & 0.028 $\pm$ 0.025 & 81.9 $\pm$ 48.0 & 0.031 $\pm$ 0.025 & 96.3 $\pm$ 15.2 \\
    3\_1 & July 12-13 Part 1 & 0.137 $\pm$ 0.060 & 151.6 $\pm$ 12.3 & 0.049 $\pm$ 0.058 & 204.5 $\pm$ 68.1 & 0.064 $\pm$ 0.059 & 93.0 $\pm$ 17.3 \\
    3\_2 & July 12-13 Part 2 & 0.078 $\pm$ 0.032 & 158.8 $\pm$ 11.7 & 0.047 $\pm$ 0.032 & 347.2 $\pm$ 37.0 & 0.062 $\pm$ 0.032 & 59.4 $\pm$ 9.8 \\
    3\_3 & July 12-13 Part 3 & 0.066 $\pm$ 0.030 & 120.5 $\pm$ 12.9 & 0.057 $\pm$ 0.030 & 78.9 $\pm$ 29.2 & 0.050 $\pm$ 0.030 & 93.6 $\pm$ 11.2 \\
    4 & July 13-14 & 0.049 $\pm$ 0.058 & 65.6 $\pm$ 32.9 & 0.044 $\pm$ 0.058 & 292.0 $\pm$ 71.4 & 0.050 $\pm$ 0.057 & 119.1 $\pm$ 20.8 \\
    \hline
    \end{tabular}
    
\end{table*}

\begin{table*}
    \centering
    \caption{Fitting parameters for the scattering-angle distributions (\enquote{modulation curves}) of \xlcp{} background observations during the 2024 flight, matching Figure~\ref{fig:Background Stokes parameters}. Parameters $A$ and $\phi$ are the amplitude and phase of three fitted harmonics, corresponding to the 180$^{\circ}$, 360$^{\circ}$ and 120$^{\circ}$ components described in~\protect\cite{AoyagiSystematicEffects}. By definition, amplitudes are positive definite. Phase angles are "random", indicating no favored direction or asymmetry in the background events. The uncertainties reported here are fitting errors only and should not be taken as true measurement errors: they are included here to indicate the phenomenological similarity between the different datasets. The limitations of the modulation-fitting approach underline the benefit of a Bayesian analysis for determining the polarization parameters PD and PA and their associated errors (Bayesian credible regions).}
    \label{tab:Fitting parameters - background}
    \begin{tabular}{|c|c|c|c|c|c|c|c|}
    \hline
    \textbf{Source} & \textbf{Date} & \textbf{$A_{180}$} & \textbf{$\phi_{180} (^{\circ})$} & \textbf{$A_{360}$} & \textbf{$\phi_{360} (^{\circ})$} & \textbf{$A_{120}$} & \textbf{$\phi_{120} (^{\circ})$} \\
    \hline
    \cyg{} & July 10-11 & 0.008 $\pm$ 0.023 & 114.5 $\pm$ 70.8 & 0.017 $\pm$ 0.023 & 119.5 $\pm$ 75.6 & 0.012 $\pm$ 0.023 & 20.5 $\pm$ 33.6 \\
    Crab & July 11 & 0.053 $\pm$ 0.033 & 93.6 $\pm$ 17.3 & 0.057 $\pm$ 0.033 & 81.9 $\pm$ 31.8 & 0.032 $\pm$ 0.033 & 36.1 $\pm$ 19.1 \\
    \cyg{} & July 11-12 Part 1 & 0.022 $\pm$ 0.038 & 118.5 $\pm$ 48.3 & 0.097 $\pm$ 0.039 & 300.6 $\pm$ 22.5 & 0.031 $\pm$ 0.038 & 63.5 $\pm$ 23.1 \\
    \cyg{} & July 11-12 Part 2 & 0.026 $\pm$ 0.033 & 158.1 $\pm$ 34.4 & 0.016 $\pm$ 0.033 & 198.7 $\pm$ 111.3 & 0.042 $\pm$ 0.033 & 27.9 $\pm$ 14.7 \\
    \cyg{} & July 11-12 Part 3 & 0.072 $\pm$ 0.032 & 151.8 $\pm$ 13.0 & 0.042 $\pm$ 0.033 & 18.2 $\pm$ 39.6 & 0.046 $\pm$ 0.032 & 43.7 $\pm$ 13.5 \\
    Crab & July 12 & 0.062 $\pm$ 0.035 & 86.3 $\pm$ 16.2 & 0.034 $\pm$ 0.035 & 138.8 $\pm$ 58.2 & 0.025 $\pm$ 0.035 & 19.4 $\pm$ 24.9 \\
    \cyg{} & July 12-13 Part 1 & 0.071 $\pm$ 0.042 & 75.2 $\pm$ 17.2 & 0.012 $\pm$ 0.041 & 356.1 $\pm$ 199.3 & 0.082 $\pm$ 0.042 & 48.4 $\pm$ 9.9 \\
    \cyg{} & July 12-13 Part 2 & 0.045 $\pm$ 0.045 & 6.9 $\pm$ 29.7 & 0.033 $\pm$ 0.045 & 77.8 $\pm$ 69.9 & 0.101 $\pm$ 0.046 & 102.4 $\pm$ 8.6 \\
    \cyg{} & July 12-13 Part 3 & 0.064 $\pm$ 0.031 & 59.7 $\pm$ 14.0 & 0.078 $\pm$ 0.031 & 117.9 $\pm$ 22.9 & 0.024 $\pm$ 0.031 & 37.6 $\pm$ 23.6 \\
    Crab & July 13 & 0.067 $\pm$ 0.029 & 154.5 $\pm$ 12.7 & 0.029 $\pm$ 0.030 & 142.9 $\pm$ 51.7 & 0.047 $\pm$ 0.030 & 34.3 $\pm$ 11.9 \\
    \cyg{} & July 13-14 & 0.021 $\pm$ 0.039 & 86.4 $\pm$ 50.8 & 0.058 $\pm$ 0.040 & 103.4 $\pm$ 37.9 & 0.050 $\pm$ 0.039 & 58.2 $\pm$ 15.3 \\
    \hline
    \end{tabular}
    
\end{table*}




\bibliography{cygvar_n2}{}
\bibliographystyle{aasjournalv7}

\end{document}